\documentclass[aps,prb,reprint,groupedaddress,showpacs]{revtex4-1}

\usepackage{color,graphicx}
\usepackage{amsmath,amsthm,amssymb,bm}
\usepackage{ulem}
\newcommand{\bv}[1]{{\bm #1}}

\newcommand{\mb}{\mu_{\text M}}
\newcommand{\ms}{\mu_{\text S}}
\newcommand{\mk}{\mu_{\text K}}
\newcommand{\Ft}{F_\text{T}}
\newcommand{\Fn}{F_\text{N}}
\newcommand{\vu}{\bv{u}}
\newcommand{\vnabla}{\bm{\nabla}}
\newcommand{\vs}{\bm{\sigma}}
\newcommand{\vse}{\bm{\sigma}^\text{(el)}}

\newcommand{\ab}{{\alpha\beta}}
\newcommand{\ep}{\epsilon}

\newcommand{\Pe}{P_\text{ext}}

\newcommand{\tx}{\tilde{x}}
\newcommand{\ttt}{\tilde{t}}

\newcommand{\tPe}{\tilde{P}_\text{ext}}

\newcommand{\tFt}{\tilde{F}_\text{T}} 
\newcommand{\tFn}{\tilde{F}_\text{N}}
\newcommand{\tV}{\tilde{V}}
\newcommand{\tv}{\tilde{v}}
\newcommand{\tu}{\tilde{u}}
\newcommand{\tua}{\tilde{u}_\text{a}}
\newcommand{\tU}{\tilde{U}}
\newcommand{\tll}{\tilde{\ell}}
\newcommand{\tlc}{\tilde{\ell}_\text{c}}
\newcommand{\tlsc}{\tilde{\ell}_\text{sc}}
\newcommand{\tL}{\tilde{L}}
\newcommand{\tH}{\tilde{H}}
\newcommand{\tlh}{\tilde{h}}
\newcommand{\tr}{\tilde{r}}
\newcommand{\tO}{\tilde{\omega}}
\newcommand{\tvc}{\tilde{v}_\text{c}}
\newcommand{\ts}{\tilde{\sigma}}
\newcommand{\tE}{\tilde{E}}

\newcommand{\teta}{\tilde{\eta}}
\newcommand{\tetat}{\tilde{\eta}_t}
\newcommand{\tld}{d}
\newcommand{\ttheta}{\alpha}
\newcommand{\tlp}{\tilde{p}}

\begin{document}


\title{Systematic Breakdown of Amontons' Law of Friction for an Elastic Object Locally Obeying Amontons' Law}


\author{Michio Otsuki* and Hiroshi Matsukawa}
\affiliation{Department of Physics and Mathematics, Aoyama Gakuin University, 5-10-1 Fuchinobe, Sagamihara 252-5258, Japan}


\begin{abstract}
In many sliding systems consisting of solid object on a solid substrate  under dry condition, the friction force does not depend on the apparent contact area and  is proportional to the loading force.
This behaviour is called  Amontons' law and indicates  that  the friction coefficient, or the ratio of the friction force to the loading force,  is constant.
Here, however, using numerical and analytical methods, 
we show that Amontons' law breaks down systematically under certain conditions 
for an elastic object experiencing a friction force that locally obeys 
Amontons' law.  
The  macroscopic static friction coefficient, which corresponds to the onset of bulk sliding of the object, decreases as pressure or system length increases.
This decrease results from precursor slips before the onset of bulk sliding,
 and is consistent with the results of certain previous experiments.
The mechanisms for these behaviours are clarified.
These results will provide  new insight into controlling friction.

\vspace{3em}

Correspondence and requests for materials should be addressed to M.O.(otsuki@phys.aoyama.ac.jp)

\end{abstract}


\maketitle

%

When we apply a shear force to a solid object on a solid substrate 
to start a sliding motion, the shear force must be greater than the maximum static friction force.
When the object is sliding, the kinetic friction force applies.
Friction plays an important role  in various  phenomena ranging from those at the nanometre scale to earthquakes;
the phenomenon of friction has been investigated since ancient times~\cite{Bauden,Robinowicz,Persson,Popov,Baumberger06}.
From the engineering point of view, friction is required to be small in certain cases and large  in other cases.
The control of friction is one of the key factors towards achieving improvements in green technology  and  nanotechnology.
In the 15th century da Vinci discovered  that the friction force is  proportional to the applied loading force and is independent of the apparent contact area between two solid surfaces~\cite{Bauden,Robinowicz,Persson,Popov,Baumberger06}.
This behaviour of friction was rediscovered by Amontons approximately 200 years later;
this law is now called
Amontons'  law of friction and holds for various systems at first approximation~\cite{Bauden,Robinowicz,Persson,Popov,Baumberger06}.
The ratio of the friction force to the loading force is called the friction coefficient, and according to Amontons' law, this coefficient does not 
depend on the loading force or the apparent contact area.
The mechanism of friction was explained in the mid-twentieth century by Bowden and Tabor\cite{Bauden}.
For actual  solid surfaces in contact  with each other, because of 
surface roughness, only a tiny fraction of the surfaces form junctions, the 
so-called real contact points.
Amontons' law is explained as resulting from the increase in the total area of real contact points, that is, the real contact area, 
in proportion to the loading force and the constant binding energy per unit real contact area~\cite{Bauden,Robinowicz,Persson,Popov,Baumberger06,Archard,Diertich}. 
However, the mechanism and the validity of  Amontons' law are still discussed  actively~\cite{Robinowicz,Persson,Popov,Baumberger06,Archard,Diertich,Communou,Gong,Bouissou98,Ben-David11,Gao,Musser}.
%

Another interesting problem concerning sliding friction is the question of how  a macroscopic object begins to slide.
Usually, it is considered that a shear force smaller than the maximum static friction force does not induce any slip motions.
However, recent measurements of the instantaneous local real contact area density of poly-(methyl methacrylate) (PMMA) show that precursors appear as  local slips at the interface under shear forces well below the maximum static friction force, 
and this force  corresponds to the onset of  bulk sliding~\cite{Rubinstein04,Rubinstein07,Ben-David10,Maegawa10}.
If the shear force  is applied slowly from the trailing edge of the sample,  a discrete propagation sequence for  the local slip appears~\cite{Rubinstein07}.
Each front of the slip starts from the trailing edge and stops after  propagating a certain length greater than that of the previous front.
When the slip front reaches the leading edge, bulk sliding occurs.
Similar behaviour is observed  in numerical studies based on 1D~\cite{Braun09,Maegawa10, Scheibert10,Amundsen} and  2D~\cite{Tromborg11} spring-block models.
The variation in the front velocity and the front velocity's
 dependence  on the stress distribution have been observed in experiments~\cite{Rubinstein04,Ben-David10} and have been examined using numerical approaches~\cite{Braun09, Tromborg11, Kammer}.
A slow precursor with a finite front velocity has been investigated with a 1D continuum model~\cite{Bouchbinder,Sinai}.
The precursor dynamics are expected to relate to the inhomogeneity of the system and are expected to be important to the understanding of friction. 
In fact, ref.~\onlinecite{Ben-David11} reported that the macroscopic static friction coefficient $\mb$, which corresponds to the maximum static friction force, varies with the experimental loading configuration, and this variation 
is related to the precursor dynamics.
Similar behaviour for  the  maximum static friction force has also 
been  investigated  with a 1D spring-block model~\cite{Ukbakh}.
Despite these studies, 
however, the mechanism of the precursor and its relation to the maximum static friction force have not been clarified theoretically.  

In this study, using both numerical and analytical methods, we investigate the precursor dynamics and the relation of these dynamics to 
the macroscopic static friction coefficient $\mb$ of an elastic object in contact with a rigid substrate,
and we show that $\mb$ decreases as pressure or system length increases.
This behaviour indicates the systematic breakdown of Amontons' law.
In the present system, the elastic object is subject to viscous damping and a friction force that obeys Amontons' law locally; however, the law 
breaks down for the overall system.
The breakdown of the law  results from a quasi-static precursor appearing 
in the propagation of local slips before the onset of bulk sliding and 
from the transition of the quasi-static precursor to a dynamic precursor at a certain critical precursor length.
For sufficiently large or small pressures or  system lengths,  
$\mb$ becomes nearly constant and Amontons' law approximately holds.
The relation of $\mb$ to the critical precursor length and the mechanism of the breakdown are clarified analytically with a 1D effective model.
These behaviours arise from  the distribution of pressure 
on the bottom of the object, which results from the torque induced by the shear force, and from the competition between the viscous damping and the velocity-weakening friction.
Consequently, the behaviours do not depend on the details of the system.
It is known that Amontons' law does not apply under  some conditions, 
such as in the presence of strong adhesion, in the absence of multiple 
real contact points, with nonlinear dependence of the real contact area on the applied load, with surface detachment, with a change in surface conditions, and for  gels~\cite{Robinowicz,Persson,Popov,Archard,Communou,Gong}.
However, the mechanism of the breakdown of Amontons' law discussed 
in this paper is different from those known previously.
In addition, a theoretical prediction of the behaviour of the friction coefficient is given for the first time.
These results  will provide new approaches to control friction.

\textbf{Results}

\begin{figure}
\begin{center}
\includegraphics[width=0.9\linewidth]{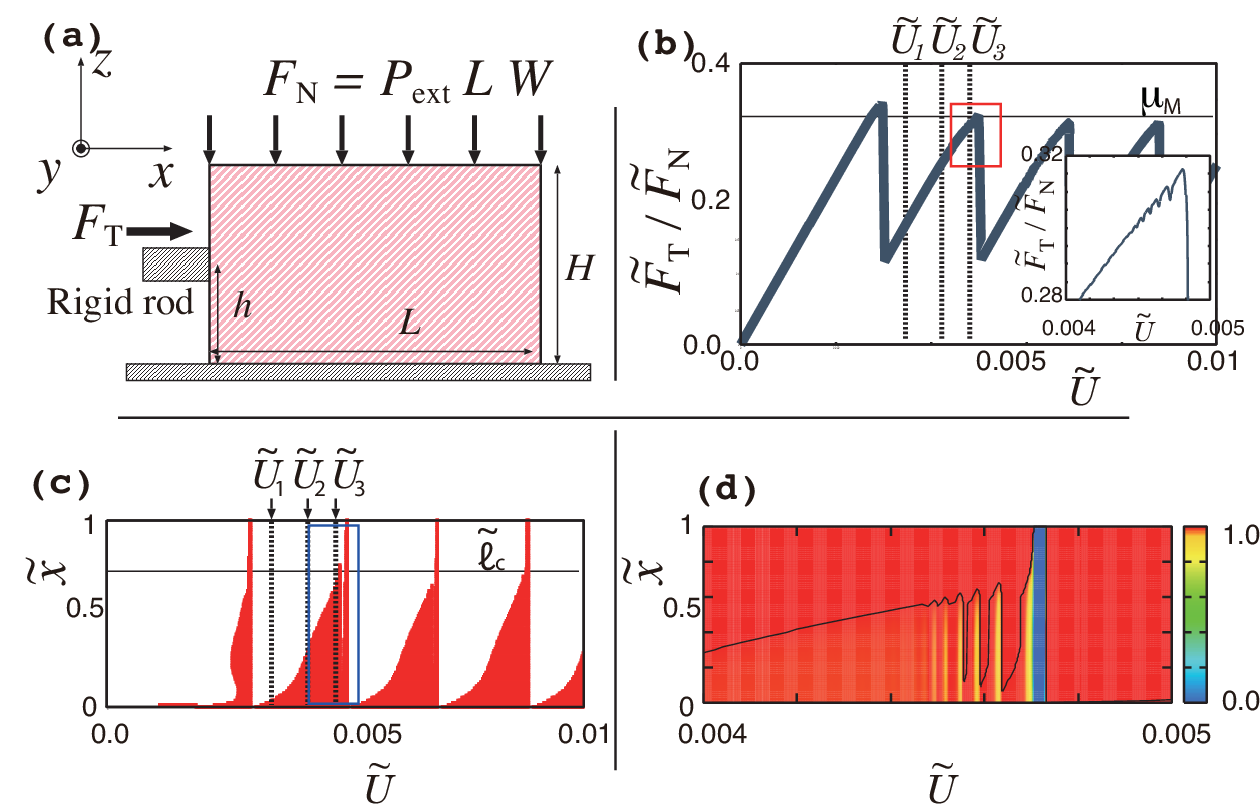}
\caption{(a) An elastic object under uniform pressure $\Pe = \Fn / (LW)$ on a rigid substrate is pushed  at height $h$ by a rigid rod with shear force $\Ft$.  
(b) The ratio $\tFt/\tFn$ as a function of $\tU \equiv\tV  \ttt$.  
The horizontal line indicates the value of  $\mb$.
The inset shows an enlarged view  of the box indicated by the red outlines.
$($c$)$ The local slip region is shown in red in the $\tU$-$\tx$ plane.  
The horizontal line indicates the critical  length of the quasi-static precursor $\tlc$. 
(d) The normalised  instantaneous local density of the real contact area in the region of the box indicated by blue outlines in (c).
The thin line indicates  the position of the precursor front.
The results are obtained for $\tL=1.0,  \tPe=0.003$ and parameter set (A).
}
\label{fig:the system}
\end{center}
\vspace{-20pt}
\end{figure}

\textbf{Finite element method (FEM) calculation.}
We first analyse the friction behaviour for an  elastic block with
 length $L$, width $W$, and height $H$ along the $x$-, $y$-, and $z$-axes, respectively, on a rigid substrate  (see Fig.~\ref{fig:the system}(a))
 using a 2D finite element method (FEM).
We assume that the block has viscous damping proportional to the strain rate 
and has friction stress  proportional to the pressure at the interface; that is, the local friction force obeys Amontons' law.
Note that the resolution of the models employed in this work is larger than the mean distance between the real contact points between
the two surfaces and smaller than the length scale of the variation of the stresses discussed below.
The validity of the law on this length scale for PMMA is supported by the proportionality of the real contact area to the loading pressure~\cite{Diertich}, with the assumption of a constant binding energy per unit real contact area,
 and the law's validity is consistent  with the friction experiments
in which the size of the apparent contact area is on the order of  
1${\rm mm}^2$ in a multiple real contact point configuration~\cite{Archard}.
The local friction coefficient $\mu\left( \dot{u}_x\right )$ is given by the local static friction coefficient $\ms$ when the local slip velocity $\dot{u}_x$ is equal to zero, and it decreases linearly from $\ms$ to the local kinetic friction coefficient $\mk$ 
with increasing $\dot{u}_x$ until the local friction coefficient finally 
equals $\mk$ at velocities greater than characteristic velocity $v_\text{c}$.
The plane stress condition is assumed  along the $y$-axis.
We apply a uniform pressure $\Pe = \Fn / (LW)$ to the top surface, where $\Fn$ denotes the loading force.
The shear force $\Ft$  
is applied from the trailing edge  at height $h$ by means of a rigid rod with a width of $0.1H$.
The rod moves with a constant velocity $V$ that is sufficiently slow.
We set $W=1$ and $h=H/2$.
The results shown here do not depend on the details of the system parameters.
However, for the quantitative calculations, we employ two parameter sets;
parameter set (A)  simulates a model viscoelastic material, and (B) simulates PMMA.  
Hereafter, we employ quantities normalised 
by the mass density, Young's modulus, and viscosity,
and these quantities are expressed with a tilde.
The following results are obtained for 
$d \equiv \tilde{H}/\tilde{L}=0.5$.
(See Methods for more details regarding the FEM calculation, the velocity dependence of the local friction coefficient, and the parameters.)

Figure ~\ref{fig:the system}(b) shows  the ratio $\tFt/\tFn$ as a function of  the displacement of the rod $\tU \equiv \tV \ttt$, where $\ttt$ denotes time.
The ratio repeatedly shows a nearly linear increase and a periodic 
large drop after the first drop.
In this work, we focus on the periodic regime.
The periodic behaviour  corresponds to a periodic stick-slip motion.
The large drop is caused  by a large slip accompanied by bulk sliding. 
The value of $\tFt$ at the peak of  the $\tFt/\tFn$ curve
is the maximum static friction force.
The inset shows an enlarged view  of the box indicated by the red outline.
A sequence of small drops in  $\tFt/\tFn$ is clearly observed 
before the onset of bulk sliding at $\tU \simeq 0.0047$.
As discussed below, each small drop is accompanied by a  rapid local slip.
This behaviour has previously been observed in certain 
experiments~\cite{Rubinstein07} and numerical studies~\cite{Braun09,Maegawa10, Scheibert10,Amundsen,Tromborg11}.
Figure ~\ref{fig:the system}$($c$)$ shows the local slip region, where the slip velocity at the interface has a finite value, in the $\tU$-$\tx$ plane.
Here, $\tx$ denotes the position at the bottom of the object.
At values well below the maximum static friction force, 
the shear force $\tFt$ induces a slow precursor slip from the trailing edge, and this precursor slip is called the {\it quasi-static precursor}.
The velocities of the local slip and the front 
of this precursor are proportional to the driving velocity,
 and they are vanishingly small.
The precursor length $\tll$ increases with increasing $\tU$.
When $\tll$ reaches a critical length $\tlc$, the quasi-static precursor becomes unstable and transforms to  a {\it leading rapid precursor} with large slip and front velocities.
The leading rapid precursor nucleates near the trailing edge.
The front of the leading rapid precursor begins with a velocity close to that of sound and in a manner similar to the supershear rupture observed in experiments~\cite{Ben-David11,Ben-David10}.  
Subsequently, the front velocity reduces  and approaches the Rayleigh wave velocity.
When the front enters the leading edge, bulk sliding occurs, and $\Ft$ shows a large drop in value.

Figure~\ref{fig:the system}(d) shows the normalised instantaneous  local density of the real contact area  in the region of the box indicated in (c). 
The local density is calculated from the local slip velocity and decreases with increasing velocity. (See Methods for this calculation.)
Before the onset of bulk sliding at $\tU \simeq 0.0047$, a discrete sequence of rapid  precursors, similar to that observed in experiments~\cite{Rubinstein07}, appears at $\tU \simeq 0.0045$, and this set of precursors causes the sequence of small drops in  $\tFt/\tFn$ 
observed in the inset of (b).
This type of precursor is called the {\it bounded rapid precursor}.
Each precursor nucleates in the region $\tilde{x} \le \tilde{\ell}$ with a front velocity that is close to the velocity of sound and is independent of the driving velocity and nucleates in a manner similar   to the leading rapid precursor.    
Subsequently, the front decelerates and stops after propagating a certain length.
The propagation length of the precursor front increases with $\tU$.
This increase is also observed in experiments~\cite{Rubinstein07}.
The local slip velocity  of this precursor is considerably  larger than that of the quasi-static precursor  shown in (c); however,  this slip
velocity decreases with decreasing driving velocity
in contrast  to the behaviour of the slip velocity for the
leading rapid  precursor. The quasi-static precursor observed in (c) almost disappears for the real contact area density shown in (d) because of  its vanishingly small slip velocity. The disappearance of the quasi-static precursor is also consistent with the results of certain experiments~\cite{Rubinstein07}.

\begin{figure}
\begin{center}
\includegraphics[width=0.7\linewidth]{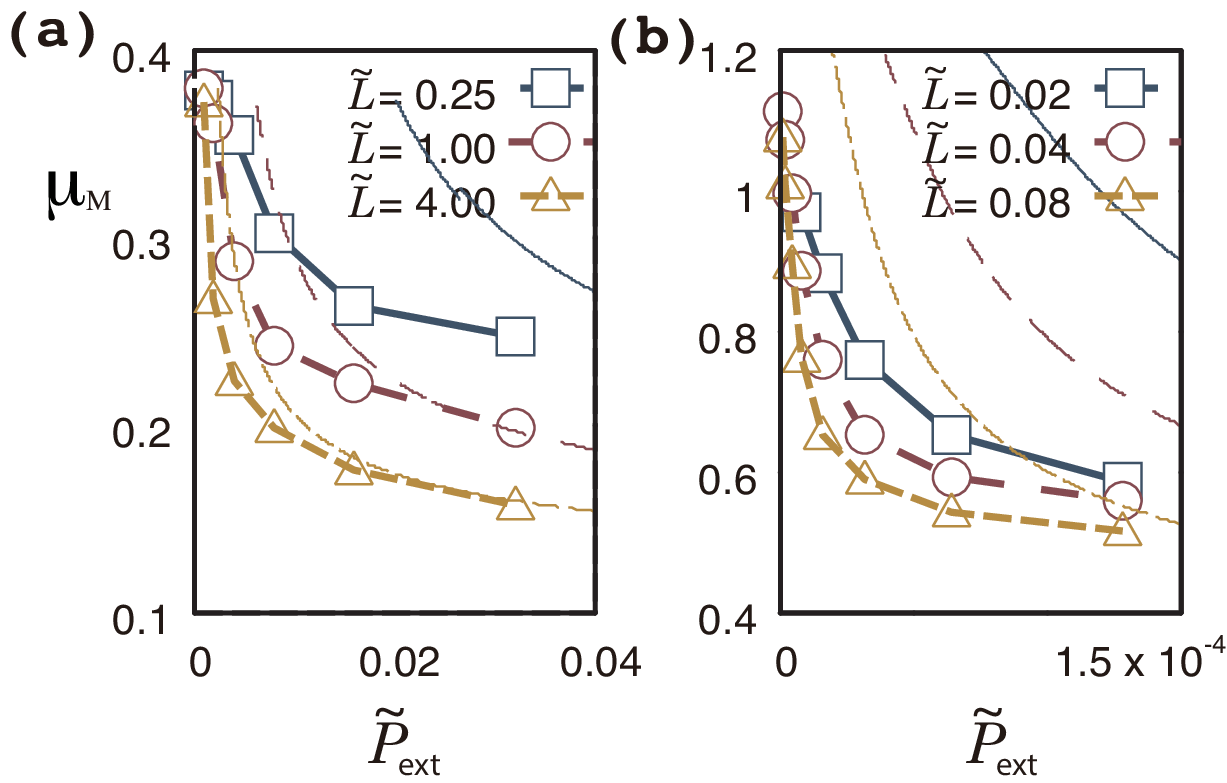}
\caption{
Macroscopic static friction coefficient $\mb$ as a function of $\tPe$  for parameter sets (A) (a) and (B) (b). The lines with symbols indicate the results of the FEM calculation.
Thin lines indicate analytical results based on the  1D effective model, where we set $\ttheta=0.2$.
Lines of the same colour correspond to the same value of $\tL$.  
}
\label{fig:myu}
\end{center}
\end{figure}

\begin{figure}[t!]
\begin{center}
\includegraphics[width=0.9\linewidth]{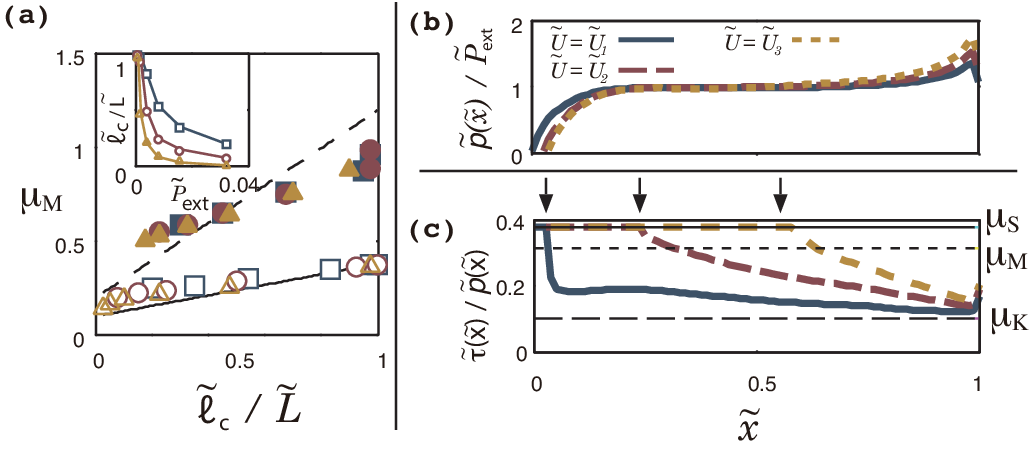}
\caption{(a) 
Macroscopic static friction coefficient $\mb$ as a function of $\tlc / \tL$ for various values of $\tPe$ and $\tL$.
The open symbols  indicate results for  the parameter set (A) with $\tL=0.25$ ($\square$), $1.0$ ($\circ$), and $4.0$ ($\triangle$) and the filled symbols 
indicate those for (B) with $\tL=0.02$ ($\blacksquare$), $0.04$ ($\bullet$), and $0.08$ ($\blacktriangle$).
The lines show the theoretical result as obtained with equation~\eqref{eq:mu}.
The inset shows the $\tPe$ dependence of  $\tlc / \tL$ for parameter set (A).
(b) The normalised pressure $\tlp(\tx)/\tPe $ and (c) the ratio of the shear stress to the pressure $ \tilde{\tau}(\tx) / \tlp(\tx)$ at the interface for the magnitudes of  $\tU$ indicated by arrows in Figs.~\ref{fig:the system}(b) and ~\ref{fig:the system}(c).
The arrows in (c) indicate the positions of the precursor front.
The three horizontal lines indicate $\ms, \mb$, and $\mk$ in order from the top of the panel.
The parameters are the same as those for Figs.~\ref{fig:the system}(b-d).
}
\label{fig:myu vs lc}
\end{center}
\vspace{-20pt}
\end{figure}

We define the macroscopic static friction coefficient $\mb$ to be the peak value of $\tFt/\tFn$ in Fig.~\ref{fig:the system}(b).
The pressure ($\tPe$) dependence of $\mb$  is shown for various values of the system length $\tL$ in Figs.~\ref{fig:myu}(a) and ~\ref{fig:myu}(b) for parameter sets (A) and (B), respectively.
Clear decreases in $\mb$ are observed with increasing $\tPe$ or $\tL$ values.
It can be shown that $\mb$ also depends on the apparent contact area.
These behaviours indicate the breakdown of Amontons' law and  the extensive property of the friction force.
This pressure dependence is consistent with the results of
experiments conducted with PMMA~\cite{Bouissou98,Ben-David11}.
The magnitudes of the normalised parameters, $\tPe$ and $\tL$, depend on the mass density, Young's modulus, 
and the viscosity coefficient.
Therefore, in contrast to the general belief  that the friction coefficient depends mainly on the surface properties, $\mb$ also depends on these bulk 
material parameters, as noted in ref.~\onlinecite{Ukbakh}.
The variation in $\mb$  results from the formation of the precursors before the onset of  bulk sliding.  
Figure~\ref{fig:myu vs lc}(a) shows that $\mb$ is scaled by the critical  length of the quasi-static precursor normalised by the system length, $\tlc / \tL$, for each parameter set.
The dependence of $\tlc / \tL$ on $\tPe$, shown in the inset,
with this scaling of $\mb$ on $\tlc / \tL$ indicates the $\tPe$ dependence of $\mb$.

Figure~\ref{fig:myu vs lc}(b) shows the pressure at the interface $\tlp(\tx)$ for the magnitudes of $\tU$ indicated by  arrows in Figs.~\ref{fig:the system}(b, c).
The applied pressure at the top surface is uniform; however, $\tlp(\tx)$  
is not uniform and increases with $\tx$.
Similar pressure distributions are observed in experiments~\cite{Ben-David11, Ben-David10} and result from the torque induced by the shear force $\tFt$~\cite{Scheibert10}.
The low pressure  causes the local maximum static friction force to be small.
As a result, the quasi-static precursor starts from  the trailing edge at $\tx=0$.
Figure~\ref{fig:myu vs lc}(c)  shows the ratio of the shear stress  at the interface $\tilde{\tau}(\tx)$ to the pressure, $ \tilde{\tau}(\tx) / \tlp(\tx)$.
Immediately after bulk sliding stops, this ratio approximately equals 
the value of $\mk$ (dashed line) for the entire interface because of the  large local slip velocity accompanied by the bulk sliding and the finite relaxation time of the local stress.
The magnitude of  $\tilde{\tau}(\tx)$ is equal to the local friction stress $\mk \tlp(\tx)$ at the instant of the vanishing acceleration of the rapid local slip, and this magnitude has almost no change during the deceleration and after stoppage because of the finite relaxation time.
However, in the region of length $\tilde{\ell}$ where the quasi-static precursor front has passed,  $ \tilde{\tau}(\tx) / \tlp(\tx)$ equals $\ms$ (straight line) because $\tilde{\tau}(\tx)$ is given by  the local friction stress for vanishing  velocity $\ms\tlp(\tx)$.
When $\tilde{\ell}$ reaches the critical length $\tlc$, the quasi-static precursor transforms to the leading rapid precursor.
Subsequently, the front of this rapid precursor enters the leading edge of the system quickly and bulk sliding occurs.
The stress distribution has almost no change during the propagation of this front  because of the short duration of the propagation.
As a result,  $\mb$ is   determined by  $\tlc$, and this relation is shown analytically below.
Note that  the stress relaxes slightly after the appearance of each of the  bounded rapid precursors, and the stress recovers its original value 
quickly through 
the following quasi-static precursor,  as shown in Fig.~\ref{fig:the system}(d).
Figure~\ref{fig:myu vs lc}(c) also shows that $\tilde{\tau}(\tx) / \tlp(\tx)$ can be larger than the macroscopic static friction coefficient $\mb$ without precipitating any local slip, as was observed in experiments~\cite{Ben-David11, Ben-David10}.

\textbf{Analysis based on a 1D effective model.}
To analyse the abovementioned numerical results, 
we employ a 1D  effective model in which the degrees of freedom of the elastic object along the $z$-axis are neglected. 
The equation of motion of the model is expressed by
\begin{equation}
\frac{\partial^2 \tu(\tx, \ttt) }{\partial \ttt^2} 
= \frac{\partial\ts_{xx}(\tx, \ttt)}{\partial \tx}  
+  \frac{\ts_{xz}(\tx, \ttt)  - \mu(\dot{\tilde{u}})\tlp (\tx)}{\ttheta\tH}.
\label{eq:evol2}
\end{equation}
Here,
$\tu(\tx,\ttt)$ denotes the displacement along the $x$-axis at the interface, $\mu(\dot{\tilde{u}})\tlp (\tx)$ is the friction stress, 
 and $\ttheta\tilde{H}$ denotes the characteristic length of the variation in the $(xz)$ component of the stress. 
We employ $\ttheta$ as a fitting parameter. 
The normal and the shear stresses are respectively given by
$
\ts_{xx} ={\tE}_1 {\partial\tu}/{\partial \tx} 
+ \tetat {\partial^2\tu}/{\partial \tx\partial \ttt}$, 
and 
$\ts_{xz} =   {\tE}_2 ({\tU - \tu})/{\tlh} +
 (\tV - {\partial\tu}/{\partial \ttt}) /{2\tlh},
$
where $\teta_t$ denotes the effective viscosity and ${\tE}_1$ and  ${\tE}_2$ represent effective elastic constants (see Methods for these parameters and the boundary condition.)
We assume that 
the pressure at the interface is given by $\tlp (\tx)= 2\tPe\tx/\tL$, which simulates the FEM result shown in Fig.~\ref{fig:myu vs lc}(b).
Hereafter, we set the origin of  $\tU$ to be the position of the pushing rod just after bulk sliding stops.
The adiabatic solution of equation~\eqref{eq:evol2} with a precursor of length $\tll$ is obtained analytically,
and the solution gives 
$\tll \simeq \tE_2 \tL\tU/\{(\ms - \mk) \tlh \tPe \}$ 
for  $\tll/\tL \ll 1$ (see Supplementary Information).
Hence, $\tll$ increases adiabatically with 
an adiabatic increase in $\tU$.
This adiabatic increase in length indicates the precursor is quasi-static.
A similar relation between the precursor length and the shear force is obtained with a 1D spring-block model; however, in this case the precursor is not adiabatic~\cite{Amundsen}.
When $\tll$ reaches a critical length $\tlc$, the quasi-static precursor becomes unstable and transforms to the leading rapid precursor.
The leading rapid precursor leads to bulk sliding.
Substituting the adiabatic equation and the relation between $\tU$ and $\tll$ into the expression for  the shear force $\tFt = \int_0^{\tL}d\tx \ts_{xz}(\tx)$ and setting $\tll=\tlc$, we obtain
\begin{equation}
\mb = \mk + (\ms - \mk)\tlc/\tL 
\label{eq:mu}
\end{equation}
for $\tlc/\tL \ll 1$.
As shown in Fig.~\ref{fig:myu vs lc}(a),
this relation agrees well with the FEM calculation, even for $\tlc/\tL\lesssim 1$.

The critical length $\tlc$ 
 is obtained by a linear stability analysis of equation~\eqref{eq:evol2}
 in the limit of vanishing  $\tV$ (see Supplementary Information).
 The analysis provides the equation for 
the $n$-th eigenvalue $\tO_n$ of the time evolution operator for the fluctuation by
\begin{align}
& \tO_n^2 \tL^2 +  \left\{ \tetat\tilde{k}_n^2 \left (\frac{\tll}{\tL} \right)^{-2} + \frac{1}{2} \left(\frac{\tr}{\tL}\right)^{-2}\right\}\tO_n \nonumber \\
&+ \tE_1 \left\{ \tilde k_n^2  \left (\frac{\tll}{\tL} \right)^{-2} + \frac{1}{\kappa^2}\left(\frac{\tr}{\tL}\right)^{-2} \right\} \nonumber \\
&-  \frac{(\ms - \mk)}{\tvc \ttheta d} \left(1- \frac{1}{\tilde{k}_n^2}\right)\tPe\tL\left(\frac{\tll}{\tL} \right)\tO_n=0 ,
\label{eq:omega}
\end{align}
where $\kappa^2 \equiv {\tE_1}/{\tE_2}$,  $\tr^2 \equiv \ttheta\tlh\tH$,  and $ \tilde{k}_n \equiv  (n +1/2)\pi$.
Here, the off-diagonal terms, which are of a higher order in $\tll$, are neglected.

The eigenvalue $\tO_n$ gives the instability condition of the adiabatic solution.
The instability results from the appearance of a positive real part for any eigenvalue.
For  small values of  $\tll/\tL$, the viscosity expressed by the second term  in equation~(\ref{eq:omega}) stabilises the adiabatic motion.
However, the velocity-weakening friction force expressed by the last term leads to instability in motion for large $\tll/\tL$.
In the present case, the oscillating fluctuation corresponding to a complex $\tO_n$ with a positive real part  never grows,
because the backward motion of the oscillation   relaxes the local shear stress, and this motion is inhibited by the friction stress.
Instead, this oscillating fluctuation yields the bounded rapid precursors observed in 
 Fig.~\ref{fig:the system}(d).
The appearance of the positive real part of $\tO_0$  at $\tll =\tlsc$, where $\tlsc$ denotes the subcritical length, yields the first bounded rapid precursor,
and the appearance of $\tO_n$ with  $n \ge 1$ is considered to yield  each of the subsequent precursors appearing at $\tll >\tlsc$.
Further increase in $\tll$ causes  $\tO_0$ to become real  and positive   at $\tlc > \tlsc$; a positive, real $\tO_0$ results in  the growing instability 
and causes the  leading rapid precursor and the subsequent bulk sliding.
The relation between  $\tlc$ and $\tPe$ is given by the instability condition, $\tO_0 > 0$.
As mentioned above, the instability is caused by the competition between the viscosity and the velocity-weakening friction force,
 which are expressed by the second and last terms  in equation~(\ref{eq:omega}), respectively.
Hence, $\tlc/\tL$ decreases as $\tPe$ or $\tL$ increases
because the last term is enhanced by $\tPe\tL$;  for $\tlc/\tL \ll 1$, $\tlc/\tL \propto (\tL\tPe)^{-1/3}$  as is easily seen from equation~\eqref{eq:omega} (also see Supplementary Information).
This result is consistent with that of the FEM calculation shown in the inset of Fig.~\ref{fig:myu vs lc}(a). 
By inserting $\tlc(\tPe,\tL)$, which is given by the abovementioned relation,  into equation~\eqref{eq:mu}, we obtain the pressure
and system length dependence 
of  $\mb$,
and $\mb$ decreases with increasing $\tPe$ or $\tL$.
The results are shown in Figs.~\ref{fig:myu}(a,b).
The analytical results agree with the FEM calculation semiquantitatively for parameter set (A) and qualitatively for parameter set (B).
The deviation of the analytical results from  the FEM calculation may arise from the absence of internal degrees
of freedom along the z-axis in the 1D effective model given by equation (1).

\textbf{Discussion}

In this work we observed three types of  precursors  prior to the occurrence of
bulk sliding: the quasi-static precursor and the bounded and the leading rapid precursors.
The latter two are seen to  correspond to those observed in experiments~\cite{Ben-David11, Rubinstein04,Rubinstein07,Ben-David10,Maegawa10}.
The quasi-static precursor is also observed in numerical studies~\cite{Bouchbinder}, but not  in experiments~\cite{Ben-David11, Rubinstein04,Rubinstein07,Ben-David10,Maegawa10}, because it  vanishes away in the  local density of the real contact area measured in experiments, as previously discussed.
Some numerical studies \cite{Maegawa10, Scheibert10,Amundsen,Tromborg11} do not report this precursor. 
The lack of this precursor is because local friction coefficient discontinuously decreases as velocity increases in these studies.
This discontinuous decrease in the local friction coefficient corresponds to  a vanishing $v_\text{c}$ and a subsequently vanishing stable region for the quasi-static precursor.
In ref.~\onlinecite{Kammer}, a model similar to the present FEM model is employed; however, 
the value of the viscosity coefficient corresponds to  $\tlc/\tL \sim 0$.
This result is also consistent with the absence of the quasi-static precursor.

In ref.~\onlinecite{Ben-David11}, it was reported that  $\mb$ varies with 
a certain acceleration length of the leading rapid precursor.
The acceleration length is difficult to define in the present work, 
and this difficulty may result from a loading configuration different from that of  the experiment.
The relation of $\mb$ with the acceleration length will be studied in future work.
However, the measured values of  $\mb$ in ref.~\onlinecite{Ben-David11} show a tendency to decrease with an increase in the loading force and depend on the stress distribution in the system.
These behaviours are consistent with those observed in the present work.

Earthquakes are among the largest friction phenomena on the earth.
Many large earthquakes are preceded by  foreshocks.
In the 2011 Tohoku-Oki earthquake with magnitude 9,  the propagation of slow slip preceding the main shock was observed over approximately 1 month~\cite{Kato}.
The behaviour is similar to the propagation of  the precursor front observed in this work.
The present results will provide new insights into earthquake behaviour.

In conclusion, using both numerical and analytical methods, we show that the macroscopic static friction coefficient $\mb$ of an elastic object  decreases with increasing pressure or system length. 
The magnitude of $\mb$ also depends on the apparent contact area and the bulk material parameters.
These behaviours indicate the systematic breakdown of Amontons' law. 
The elastic object is subject to viscous damping and a friction force that obeys Amontons' law locally; 
 however, the law undergoes breakdown as a whole.
The behaviour of  $\mb$ is consistent with the results of relevant 
experiments~\cite{Bouissou98,Ben-David11}, and it arises from the occurrence of precursor slips before bulk sliding and  from the transition from quasi-static to rapid motion of the precursors.
The linear stability analysis based on a 1D effective model gives the relation between $\mb$ and the critical length of the transition, and it clarifies the mechanism of the transition.
The transition is caused by  the pressure distribution at the bottom of the object, which  results from the torque induced by the shear force, and by the competition between the viscous damping and the velocity-weakening friction.
If the ratio of the critical length of the transition to the system length is considerably less than or approximately equal to unity, $\mb$ becomes almost constant and Amontons' law holds approximately.
These situations are caused by sufficiently large or small values of pressure or system length.
The qualitative features of these results do not depend on the details of the model for the case in which the pressure at the bottom of the object increases monotonically along the driving direction, the local friction coefficient decreases with velocity continuously, and the system has viscous damping.
According to these results, an object with greater width
or an object in which the part contacting with  the interface is divided into shorter portions along the driving direction can have a larger maximum static friction force under a constant loading force.
The present results will provide new techniques for controlling friction.

\textbf{METHODS}

\textbf{FEM calculation.}
The equation of motion of the displacement vector  of the elastic object $\vu(\bv{r},t)$ employed in the FEM calculation is expressed as $
\rho \partial^2\vu(\bv{r}, t)/\partial t^2= \vnabla \cdot \vs$, 
where $\rho$ is the mass density.   
The  stress tensor $\vs$ is composed of an elastic part $\vse$ and a dissipative part.
The dissipative component is expressed as
$\sigma_\ab^{\text{(dis)}} = \eta_1\dot{\ep}_{\ab} + \eta_2(\dot{\ep}_{xx}+\dot{\ep}_{zz})\delta_\ab$.
Here $\dot{\ep}_\ab$ denotes a component of the strain rate tensor, and $\eta_{1,2}$ denote the viscosity coefficients~\cite{Landau}.
The boundary condition  is given by 
$\sigma_{zz} = -\Pe \text{ and } \sigma_{xz} = 0$ at the top surface,  and 
$\sigma_{xx} =0 \text{ and } \sigma_{zx} = 0$  at $x=0$ and $L$ except 
for the portion in contact with the rigid rod.

In the FEM calculation, 
the object is divided into equal-sized rectangular cells.
The number of cells is $40\times 40$ or $80\times 80$.
The convergence of the results with respect to the FEM mesh size is verified.
At the beginning of the FEM simulation, we apply a uniform pressure to the top surface causing the elastic object to relax.
Subsequently, we start pushing the object at $\ttt=0$.
The results shown in the paper
are obtained for $\tilde{V}=10^{-5}$.
This value of $\tilde{V}$
corresponds to the adiabatic limit
except for the results shown in the inset of Fig.~\ref{fig:the system} (b) and in Fig.~\ref{fig:the system} (d).

\textbf{Velocity dependence of the local friction coefficient 
and the real contact area density.}
The velocity dependence of the local friction coefficient employed in this work is obtained  from the rate- and state-dependent friction law, $\mu = \mk + (\ms-\mk)\phi$ with $\dot{\phi}=(1-\phi)/\tau -|\dot{u}_x|/D$,  as discussed in ref.~\onlinecite{Popov}.
Here, $0 \le \phi \le 1$ represents the state variable and $\tau$ and $D$ denote the relaxation time and length, respectively.
The above equation yields $\mu(\dot{u}_x)$, 
as noted in the text for $v_\text{c} = D / \tau$,
in the limit of vanishing $\tau$, where $\phi = 1-|\dot{u}_x|/v_\text{c}$.
The magnitude of $\phi$ is proportional to the deviation of the local density of the real contact area from  that for $|\dot{u}_x| \ge v_\text{c}$.
Thus, we obtain the normalised instantaneous local density of the real contact area shown in Fig. \ref{fig:the system}(d). 

\textbf{Parameters.}
We introduce the normalised quantities in which 
the length, mass and time are normalised by $L_0 \equiv \eta_1 / \sqrt{\rho E}$, $m_0 \equiv \eta_1^3 / \sqrt{\rho E^3}$ and $t_0 \equiv\eta_1/E$, respectively, where $E$ represents Young's modulus.
In this paper, the normalised quantities are expressed with a tilde.
We employ two parameter sets:
parameter set (A) simulates a model viscoelastic material, 
which has 
 Poisson's ratio $\nu=0.34,  \ms=0.38, \mk=0.1, \tv_c=3.4 \times 10^{-4}$, and $\teta_2=1$;
and
 (B) simulates PMMA,
 which has
 $\nu=0.4, \ms=1.2, \mk=0.2, \tv_c=3.9 \times 10^{-7}$, and $\teta_2=1$.
The present FEM calculation shows that $\ms$ and $\mk$ correspond to the maximum and minimum values of the ratio of the local shear stress to the pressure, respectively.
Hence we estimate $\ms$ and $\mk$ for PMMA from the maximum value of the ratio in Fig.~4(c) and the minimum value of the ratio in Fig.~4(b) in ref.~\onlinecite{Ben-David11}, respectively.  
The value of $v_\text{c}$ is estimated from the values of $\ms, \mk$ and the linear fit of the steady-state velocity dependence of the rate- and 
state-dependent friction law reported  in ref.~\onlinecite{Baumberger06}.
The value of $L_0$ for PMMA is estimated as follows.  
In ref.~\onlinecite{Ben-David11}, $\mb=0.6$  for $L=0.2 \text{ m}, d=0.5$, and $P_\text{ext}=1.67 \times 10^6  \text{ Pa}$.
The FEM calculation yields $\mb=0.6$ for $\tPe \tL = 2.3 \times10^{-6} $.
Consequently, we obtain $L_0=48 \text{ m}$.
These values of the parameters for PMMA have some uncertainties.
However, the essential feature of the results obtained here is, as mentioned before, not specific to these values.
The local validity of Amontons' law for PMMA may require some discussions.
Even if Amontons' law breaks down locally, 
the behaviour of the precursor dynamics and the static friction force discussed here will be still qualitatively valid because their mechanism is general.
For metals, Amontons' law is expected to better hold locally and 
the present qualitative results will also be applicable. 

\textbf{1D effective model.}
In the 1D effective model, the effective viscosity is given by $\teta_t \equiv 1+\teta_2$ and the two elastic constants are given by  ${\tE}_1 \equiv  1 / \{(1+\nu) (1-\nu) \}$ and ${\tE}_2 \equiv 1 /\{2 (1+\nu)\}$.
The boundary condition is given by 
${\partial \tu(\tx, \ttt)}/{\partial \tx}  =0$ at $\tx=0$ and $\tL$.


\begin{acknowledgments}

The authors thank J. Fineberg, S. Maegawa, K. Nakano, M. O. Robbins, K. M. Salerno, A. Taloni  and T. Yamaguchi for valuable discussions.
This work was supported by KAKENHI (22540398) and (22740260) and  by HAITEKU
from MEXT.
\end{acknowledgments}

\textbf{Author Contributions}

M.O. performed the FEM calculations. All the authors contributed to the analysis and writing of the manuscript.

\textbf{Additional Information}

\textbf{Supplementary Information} accompanies this paper at http://www.nature.com/ scientificreports

\textbf{Competing financial interests:} The authors declare no competing financial interests.

\clearpage

\begin{center}

{\Large Systematic Breakdown of Amontons' Law of Friction in Elastic Object Subject to Amontons' Law Locally}

Michio Otsuki and Hiroshi Matsukawa\\

Department of Physics and Mathematics, Aoyama Gakuin University, 5-10-1 Fuchinobe, Sagamihara 252-5258, Japan\\
\end{center}

\begin{center}
{\large Supplementary Information }
\end{center}

{\large \textbf{Supplementary Notes} }

Here we provide the explicit forms of the adiabatic solutions of equation (1), and the relation between  $\tlc$ and $\tPe$ discussed in the text.
We also discuss the linear stability analysis in greater detail.

The adiabatic solution of equation (1) for $\tU=0$ is given as
\begin{equation*}
\tu_0(\tx) = \frac{ 2\mk\tlh\tPe  }{\tE_2 \tL}
\Big\{ \frac{ \kappa \tr}{\sinh\frac{\tL}{\kappa \tr}} \big( \cosh\frac{\tx}{\kappa \tr} - \cosh\frac{\tL-\tx}{\kappa \tr}  \big) - \tx \Big\}, 
\end{equation*}
where we set the local friction coefficient to $\mk$.
For $\tU>0$, the  precursor proceeds to $\tll$.
The adiabatic solution for $\tU>0$, $\tua (\tx)$, is given by
$\tu_1(\tx) \mbox{ for  $0 \le \tx \le \tll$ and }
\tu_0(\tx)   \mbox{ for $\tll \le \tx$}$.
%
%
In the region $\tll < \tx$, the increase in the shear stress with increasing $\tU$ is canceled  by the increase in the local friction stress, which can take a value between $\mk\tlp(\tx)$ and $\ms\tlp(\tx)$.
The solution is subsequently given by $\tu_0(\tx)$.
The solution in the region $0 \le \tx \le \tll$, where the local friction coefficient is equal to $\ms$, is 
obtained as
\begin{alignat*}{2}
&\tu_1(\tx) = \tu_0(\tx)+\tU\big( 1- \cosh \frac{\tx -\tll}{\kappa \tr}\big) \nonumber \\
&+\frac{ 2(\ms - \mk)\tlh \tPe  }{\tE_2 \tL} \Big\{ \ell \cosh \frac{\tll-\tx}{\kappa \tr} - \kappa \tr \sinh \frac{\tll-\tx}{\kappa \tr} -   \tx \Big\} .
\end{alignat*}

%

%

The linear stability of the adiabatic solution of the 1D effective model is analyzed as follows.  
Setting $\tu(\tx,\ttt)=\tu_a(\tx) + \delta\tu(\tx,\ttt)$, 
and linearizing equation (1) with respect to the fluctuation $\delta\tu(\tx,\ttt)$, we obtain its time evolution equation.
The boundary condition is given by $\frac{\partial}{\partial \tx}\delta\tu(\tx,\ttt)|_{\tx=0}=0$ and $\delta\tu(\tx>\tll,\ttt)=0$.
We express $\delta \tu(\tx,\ttt)  =   \sum_{m=0} \tilde{u}_m\exp{(\omega_m\ttt)} \cos(k_m \tx / \tll )$, where $ \tilde{k}_m =  (m +1/2)\pi$, and obtain  equation (3).

The relation between  $\tlc$ and $\tPe$ given by 
the condition that $\omega_0$ becomes real and positive is expressed as%
\begin{align}
&\tPe= \frac{\pi^2}{\pi^2-4}\frac{\tvc\ttheta\tld}{(\ms - \mk)} \frac{1}{\tlc}   \nonumber \\
&\times\bigg\{ \tetat\Big(\frac{\pi\tL}{2\tlc}\Big)^2 + \frac{1}{\ttheta d^2}  
+2\tL\sqrt{ \tE_1\Big(\frac{\pi\tL}{2\tlc}\Big)^2+  \frac{2\tE_2}{\ttheta d^2} } \bigg\} . \nonumber 
\label{eq:pe}
\end{align}

\end{document}